\documentclass[traditabstract]{aa}
\usepackage{graphicx}
\usepackage{natbib}
\usepackage{ulem}

\bibpunct{(}{)}{;}{a}{}{,} 
\newcommand{\ind}[1]{_{\mathrm{#1}}}

\begin{document}

\title{Detection of Jovian seismic waves: a new probe of its interior structure}
\author{P. Gaulme\inst{1}
          \and F.-X. Schmider\inst{2}
          \and J. Gay\inst{2}
          \and T. Guillot\inst{3}
          \and C. Jacob\inst{2}
          }

\offprints{P. Gaulme}

   \institute{$^1$ Institut d'Astrophysique Spatiale (UMR 8617), Facult\'e des Sciences d'Orsay, Universit\'e Paris Sud, B\^atiment 121, F-91405 ORSAY Cedex  \\
 $^2$  Laboratoire Fizeau, Universit\'e de Nice, CNRS-Observatoire de la C\^ote d'Azur, F-06108 Nice Cedex 2, France\\
 $^3$  Laboratoire Cassiop\'ee, Universit\'e de Nice, CNRS-Observatoire de la C\^ote d'Azur, F-06108 Nice Cedex 2, France\\
              \email{Patrick.Gaulme@ias.u-psud.fr}
        }

\titlerunning{Detection of Jupiter's global oscillations}
\authorrunning {Gaulme et al.}
\abstract
{Knowledge of Jupiter's deep interior would provide unique constraints on the formation of the Solar System. Measurement of its core mass and global composition would shed light on whether the planet formed by accretion or by direct gravitational collapse. At present, the inner structure of Jupiter is poorly constrained and seismology, which consists of identifying acoustic eigenmodes, offers a way to directly measure its deep sound speed profile, and thus its physical properties. Seismology of Jupiter has been considered since the mid 1970s, but hitherto the various attempts to detect global modes led, at best, to ambiguous results. We report the detection of global modes of Jupiter, based on radial velocity measurements performed with the SYMPA Fourier spectro-imager. The global seismic parameters that we measure include the frequency of maximum amplitude $1213\pm50\ \mu$Hz, the mean large frequency spacing between radial harmonics $155.3\pm2.2\ \mu$Hz and the mode maximum amplitude $49_{-10}^{+8}$ cm s$^{-1}$, all values that are consistent with current models of Jupiter. This result opens the way to the investigation of the inner structure of the Solar System's giant planets based on seismology techniques.
}
\keywords{Planets and satellites: interiors, Jupiter - Waves -Asteroseismology - Techniques: radial velocities - Methods: data analysis}

\maketitle
\section{Introduction}
Jovian seismology has long been considered as both a potentially powerful tool for probing the interior of Jupiter \citep{Guillot_2004} and a natural extension of helioseismology \citep{Goldreich_Keeley_1977}, because the common fluid nature of Jupiter and the Sun is expected to lead to similar oscillations and the possibility of using similar observational techniques. Theoretical works \citep{Vorontsov_1976,Bercovici_Schubert_1987} predict that Jovian global oscillations should have a frequency range of [800, 3500] $\mu$Hz with 10 to 100 cm s$^{-1}$ amplitude, values that are comparable to those of the Sun. Several attempts to observe Jovian global modes have been carried out since the mid-1980s, with thermal infrared photometry \citep{Deming_1989}, and Doppler spectrometry using both magneto-optic \citep{Schmider_1991} and Fourier transform spectrometers \citep{Mosser_1993,Mosser_2000}, respectively. Moreover, infrared observations \citep{Walter_1996,Mosser_1996} were performed to look for pressure waves excited by the impact of the comet Shoemaker-Levy 9. Infrared observations are affected by atmospheric inhomogeneities and have been unsuccessful so far. In contrast, all of the Doppler measurements exhibit an excess power of about 1 m$^2$ s$^{-2}$ between 800 and 2000 $\mu$Hz, which cannot be explained by instrumental systematics nor spurious atmospheric signals. A tentative comb-like structure with a $139\pm3\ \mu$Hz mean spacing was also identified, but the combination of a low signal-to-noise ratio (hereafter SNR), this frequency corresponding to the least common multiple of terrestrial and Jovian rotation frequencies, and the spacing being incompatible with theoretical estimates \citep{Provost_1993,Gudkova_1995,Gudkova_Zharkov_1999} implied that it was probably an artifact.
The need for specific instrumentation able to combine high spectral and spatial resolution thus emerged from the 1980 and 1990s experiments. The major difficulty in performing seismic observations of Jupiter is related to its rapid rotation (500 m s$^{-1}$ at the equator on 1 arcsec), which diminishes the instrument's velocity sensitivity and makes it extremely sensitive to pointing errors. The SYMPA instrument (\citealt{Schmider_2007}, hereafter S07) is a Fourier tachometer designed to circumvent this difficulty, whose principle is based on the spectro-imaging of the full planetary disk in a non-scanning mode, and is related to the helioseismic instruments GONG \citep{Beckers_Brown_1979} and MDI/SOHO \citep{Scherrer_1995}. It produces radial velocity maps of Jupiter's upper troposphere by measuring the Doppler shift of solar Mg lines at 517 nm that are reflected by Jupiter's clouds (S07).

Two instruments were built and simultaneously used during two observation runs in 2004 and 2005 at San Pedro Mart\'ir and Teide observatories.
The best observation sequence with SYMPA was acquired as part of the 10-day run performed at the Teide Observatory between April 2 and 12 2005. Jupiter was then in opposition with positive declination. Such favorable conditions for observations did not occur again in the northern hemisphere until January 2011. The data were found to be of much higher quality than all previous observations: at the end of the run, with a 21.5\% duty cycle, the mean noise level reached 12 cm s$^{-1}$ per sample, which was five times lower than previously achieved for Jupiter \citep{Mosser_2000}. At the time, a preliminary analysis did not yield the detection of Jovian modes (\citealt{Gaulme_2008a}, herafter G08).

Here we introduce the detailed analysis of the 2005 data obtained at Teide Observatory, where we focus on the time series with the highest SNR. A short review of the data processing explains the procedure that we followed to optimize our chance of finding out the global modes, instead of performing a global analysis as in G08. This leads to the detection of an excess power modulated by a comb structure, whose properties are compatible with Jovian modes (Sect. 2). We then explore all of the possible origins of the detected signal to conclude of the detection of Jupiter's global modes (Sect 3). Last, we conclude on the implication that this result generates about future studies of giant planets (Sect. 4).

\section{Detailed analysis of the SYMPA data}

\subsection{Data processing short review}
We review the essential steps of the whole data processing, from a single acquisition to the production of power spectra associated with a set of spherical harmonics. We refer the reader to S07 and G08 for a detailed description.

SYMPA is a Fourier tachometer based on a Mach-Zehnder (MZ) interferometer working at fixed optical path difference. It produces four interferograms of the solar spectrum reflected by Jupiter, in a 5-nm spectral range centered around the Mg triplet at 517 nm.  Each interferogram modulates the Jovian photometric figure. The differences between the two pairs of images allow us to cancel the photometric intensity, to produce two fringe patterns that are phase-shifted by about 90 degrees. This step involves the application of an algorithm that straightens up each image from geometrical distortions, then overlaps the two pairs of images. However, the accuracy of this process was insufficient to fully cancel the Jovian image in both interferograms. In particular, Jupiter's zones and belts were still slightly visible. Thus, we applied a Wiener filter in the 2D Fourier transform of both interferograms in order to substantially reduce the photometric residuals. Moreover, towards the limb, the SNR is low because of Jupiter's limb darkening, while geometrical distortions are maximum. Therefore, the velocity maps are limited to the central 75\% of the Jovian radius, applying an apodisation at the edge.

The linear combination of the two interferograms leads, after substraction of the instrumental phase, to a radial velocity map. In this map, the Jovian rotation is the major effect. The first step consists of subtracting from the velocity map the Jovian rotation plus the relative motion of Jupiter with respect to the observer. This requires us to precisely determine the inclination of Jupiter's rotation axis on the image, which is achieved within a fraction of a degree thanks to the belts and zones on Jupiter. The rotation law is then applied to the data after centering the image. The precision of the centering is one of the limiting factors of the instrument's performance.
The velocity maps were then decomposed into spherical harmonics: each map was multiplied by a set of masks, made of the projections of spherical harmonics in two dimensions. All pixels of each filtered map were then weighted with spherical harmonics to create a set of time series. The power spectrum of each time series was computed with a discrete Fourier transform.
Spherical harmonics were considered until degree $\ell = 25$, as it matches the maximum spatial resolution of SYMPA.

\begin{figure}
\includegraphics[width=9.4cm]{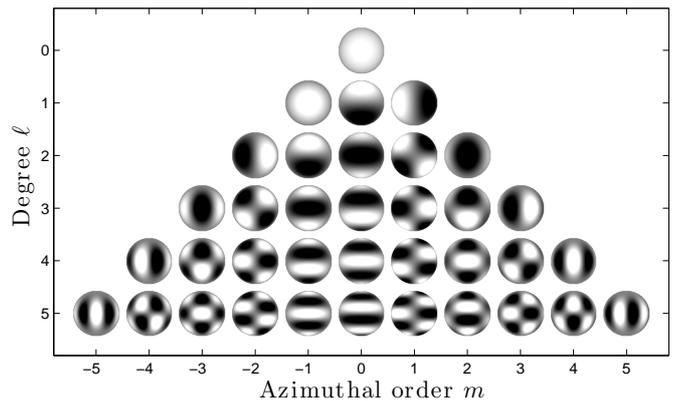}
\caption{Representation of the projected spherical harmonics, as a function of degree $\ell$ and azimuthal order $m$, until degree $\ell=5$. The masks used to filter the velocity maps are limited to 75\% of the Jovian diameter. }
\label{fig_0}
\end{figure}

\begin{table}
\caption{Coupling between the masks, with value larger than 20\%.}
\begin{tabular}{| l | l l l l l |}
\hline
$\ell=0$ & $\ell=1$  & $\ell=2$   &    $\ell=2$   & $\ell=3$  & $\ell = 3$\\
$m=0$   & $m=-1$   & $m=0$     &    $m=2$     & $m=-3$   & $ m =-1$\\
                &    98.0\% & -66.7\%   &    -79.5\%    & -40.1\%    & -33.8\%  \\
\hline
\hline
$\ell=1$ & $\ell=2$  & $\ell=3$   &    $\ell=3$   & $\ell=4$  & $\ell = 4$\\
$m=0$   & $m=-1$   & $m=0$     &    $m=2$     & $m=-3$   & $ m =-1$\\
                &    97.0\% & -72.1\%   &    -77.0\%    & -37.8\%    & -40.7\%  \\
\hline
\hline
$\ell=1$ & $\ell=2$  & $\ell=3$   &    $\ell=3$   & $\ell=4$  & $\ell = 4$\\
$m=1$   & $m=-2$   & $m=1$     &    $m=3$     & $m=-4$   & $ m =-2$\\
                &    97.0\% & -45.2\%   &    -79.8\%    & 45.4\%    & 23.7\%  \\
\hline
\hline
$\ell=2$ & $\ell=3$  & $\ell=4$   &    $\ell=4$   & $\ell=5$  & $\ell = 5$\\
$m=1$   & $m=-2$   & $m=1$     &    $m=3$     & $m=-4$   & $ m =-2$\\
                &  -96.4\% & -56.6\%    &    -78.2\%    & 43.7\%    & 31.8\%  \\
\hline
\end{tabular}
\label{tab_1}
\end{table}

\begin{figure*}[t]
\hskip 1.3 cm
\includegraphics[width=15cm]{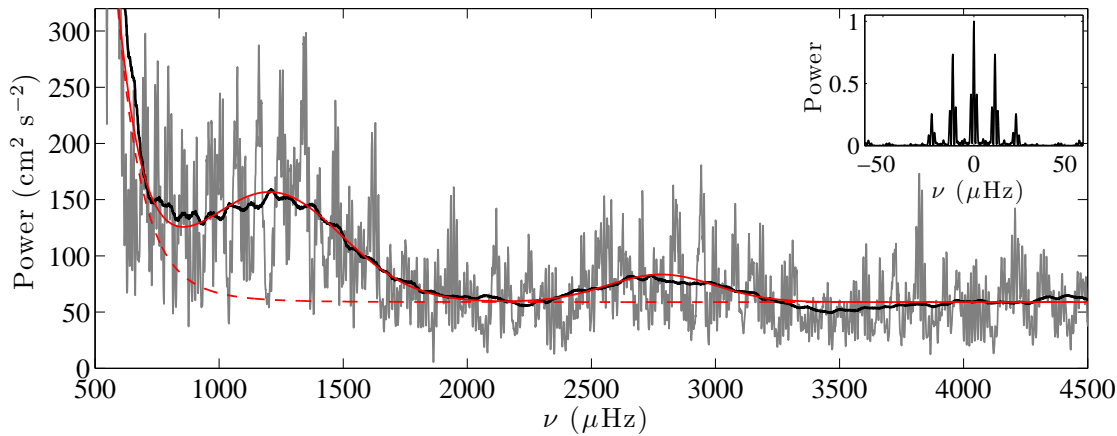}
\caption[]{Power spectrum of the time series associated with the spherical harmonics of degree l = 1. The thin grey and bold black lines represent the power spectrum smoothed with a moving average, weighted by a triangular window, with two different scales. The grey line uses a 3.1-$\mu$Hz wide window, and the black line a linearly increasing width window, from 14 to 129 µHz across the [500, 4500] $\mu$Hz frequency range. The red dashed line indicates the background level fit and the full red line the background plus excess power fit. The upper plot represents the power spectrum of the window function, normalized such that its maximum value equals 1. }
\label{fig_1}
\end{figure*}

\subsection{Data with best SNR}
The rapid global analysis of the 2005 dataset (G08) did not yield a detection of global oscillations of Jupiter because of three independent factors.

Firstly, the calculation of the power spectra discovered a spurious signal, at $5920\pm200\ \mu$Hz, which is strongly correlated with the position of Jupiter on the detector. This signal is due to the imperfect disentangling of the photometric and spectrometric contents. In particular, its amplitude is mostly a function of the fluctuations in the position of Jupiter along the $x$-axis. This dependence on the position of Jupiter ensures that the time series associated with symmetrical masks is more sensitive to the systematic errors. Moreover, an additional spurious signal was found, with a central frequency proportional to the product of the azimuthal order m times the Jovian rotation frequency. From order m = 5, this spurious signal contaminates all power spectra at frequencies higher than 800 $\mu$Hz. This artifact impedes the search for Jovian global oscillations in the power spectra associated with spherical harmonics of azimuthal order m $\leq$ 5.

The second factor is related to the blurring of the Jovian disk that is generated by both atmospheric seeing and random guiding fluctuations. The mean standard deviation of the position of Jupiter on the detector was measured to be 3.3 pixel rms, for the whole observation run, i.e, 2.1 arcsec. Since the apparent diameter of Jupiter was 44 arcsec in April 2005, only 20 elements of resolution can be considered to be independent along its diameter. Moreover, the seeing varied during the observations and could sometimes reach a value of 2 arcsec, which contributed significantly to the blurring. Because we limited the analysis to 75 \% of the Jovian radius, this blurring prevents us from searching for modes of degree higher than $\sim8$. It also signifies that the mode sensitivity decreases as the degree increases.

Thirdly, the projection of spherical harmonics in two dimensions generates a non-orthonormal base. The elimination of the external ring of 25\% of Jupiter's diameter, which contributes more to the noise than to the signal, dramatically increases the degeneracy of the projected base. In Table \ref{tab_1}, we present the coupling of these four masks with their corresponding correlated masks. It turns out that if we consider the spherical harmonics until degree l=5, only four masks remain independent: (l,m = 0,0), (l,m = 1,0), (l,m = 1,$\pm$1), and (l,m = 2,$\pm$1).
Instead of globally analyzing all of the available data, as performed in G08, the present analysis focuses on the time series with the highest SNR. The probability of finding modes in our data is maximized by dealing with: i) antisymmetric masks, ii) azimuthal orders m $<$ 5, iii) degrees as low as possible, and iv) independent masks to avoid redundancy. Therefore, we hereafter consider the two time series that are filtered with the masks associated with spherical harmonics of degree and azimuthal order: l,m = 1,0 and l,m = 1,$\pm$1. Unlike a classical helioseismology analysis \citep{Schou_1992}, we do not apply the Legendre polynomial function to each m value, but apply only the antisymmetric masks (Fig. \ref{fig_0}). Hence, the time series contains both prograde and retrograde modes when the azimuthal order is not null. This procedure allows us to eliminate most of the spurious signal that would globally affect the velocity map, as guiding error and MZ temperature variations.

\subsection{A clear excess power modulated by a comb-like structure}
We sum the information contained in these two independent time series by computing the power spectrum of the mean time series. The data set constitutes 27\,946 measurements acquired between April 2 at 23:48:14 (UT) and April 12 at 06:30:17 (UT), with a 6-s time sampling and a 21.5\% duty cycle. The power spectrum is computed with the discrete Fourier transform with an oversampling of a factor of two and a frequency resolution of 1.248 $\mu$Hz (Fig. \ref{fig_1}). We then analyze the power spectrum with methods similar to those developed for helioseismology \citep{Harvey_1985} and successfully applied to asteroseismic data from CoRoT \citep{Michel_2008} and Kepler \citep{Chaplin_2010}. As in previous Doppler measurements, the time series exhibits excess power between 800 and 2000 $\mu$Hz and a secondary excess between 2400 and 3400 $\mu$Hz. This time it is modulated by a clear comb-like structure of high SNR.

\begin{figure}[t]
\includegraphics[width=9cm]{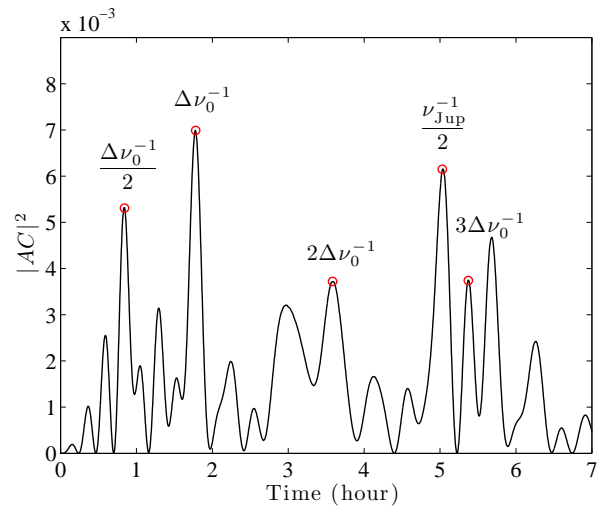}
\caption[]{Power spectrum of the [750-2050 $\mu$Hz] power spectrum of the time series. It is equivalent to the square of the envelope of the autocorrelation of the time series $|AC|^2$, and is computed as the Fourier transform of the power spectrum filtered with a tapered cosine window centered on 1400 $\mu$Hz and with 650 $\mu$Hz half-window. It is normalized such that $|AC(0)|^2 = 1$, then filtered for correlation times lower than 0.3 h. On one hand, 4 peaks, labeled (1/2,1,2,3) $\Delta\nu_0^{-1}$, are associated with the mean spacing of the modes: $\Delta\nu_0  = 155.3\pm2.2\ \mu$Hz. The corresponding period, $1.789\pm0.026$ h, matches twice the sound speed transit time between the centre and the surface of Jupiter. On the other hand, the second highest peak at 5.030 h almost matches twice the Jovian rotation period $\nu\ind{jup}$, a spacing that is compatible with the mode splitting between positive and negative azimuthal orders (e.g. l,m=1,+1 and l,m=1,-1).}
\label{fig_2}
\end{figure}

We fit the mean background with a semi-Lorentzian function and each power excess with a Gaussian function, using the maximum likelihood estimator by taking into account the nature of the statistics, a $\chi^2$ with two degrees of freedom. From this fitting, the frequencies of maximum amplitude of both excess powers are $1213\pm50$ and $2781\pm49\ \mu$Hz, with amplitudes of $9.8\pm3.0$ cm s$^{-1}$ and $5.0\pm2.3$ cm s$^{-1}$, respectively. The white noise has an amplitude of $7.7\pm1.0$ cm s$^{-1}$. To estimate the splitting frequency of the comb-like structure, we calculate the Fourier transform of the power spectrum by restricting the frequency domain to the main excess power [800, 2100] $\mu$Hz. The resulting power spectrum as a function of time (Fig. \ref{fig_2}) is equivalent to an autocorrelation of the time series. It is characterized by a main peak at 1.780 h, with second and third harmonics at 3.584 h and 5.373 h and a half harmonic at 0.845 h. A linear fitting of the four values yields a correlation time of $1.789\pm0.026$~h, corresponding to a frequency spacing of $\Delta\nu_0 = 155.3\pm2.2\ \mu$Hz. The fact that we observe a half harmonic in time rather than only integer harmonics can be due either to a combination of the window observing function with the modulation of nearby oscillation modes or to a stochastic excitation of the modes. We also detect the presence of a peak at half the Jovian rotation period as would be expected from the rotational splitting of l=1 azimuthal modes.
The \'echelle diagram associated with this mean spacing frequency $\Delta\nu_0$ (Fig. \ref{fig_3}) clearly shows two maxima in the $\sim$1000-1600 $\mu$Hz range that are separated by half of the 155 $\mu$Hz folding frequency. Furthermore, two secondary maxima of smaller amplitude mirror them in the 2500-3000 $\mu$Hz domain.

To assess the significance of the observed signal, we then apply a null hypothesis (H0) statistical test to the peaks identified in the power spectrum. We retain only peaks with a significance level of 1\% and remove all neighboring aliases caused by the Earth's rotational frequency ($\pm11.6\ \mu$Hz). This results in the identification of 22 distinct local maxima between 800 and 3100 $\mu$Hz. Both amplitude and frequencies were measured from the power spectrum rebinned over 11 points, to smooth out the daily aliases.
To estimate the true amplitude of the detected peaks, we introduced 40 pure sine curves with amplitudes corresponding to a Doppler shift of 1 m s$^{-1}$ in the time series, at frequencies away from the excess power. Because of the duty cycle and the rebinning, the sine curves appear as peaks of mean amplitude of [$12.2_{-1.0}^{+2.0}$~cm s$^{-1}$]$^2$ in the power spectrum. This scaling factor was used to infer the peak heights. If we assume these peaks are of Jovian origin, we then have to divide each peak height by a factor two, to take into account the doubling of the Doppler effect that affects the shift of reflected lines. By supposing these peaks correspond to pure sinusoidal signals, the maximum velocity for these is $49_{-10}^{+8}$~cm s$^{-1}$. Moreover, the frequency measurement of 40 artificial sine curves in the data allowed us to estimate the error in the frequency to be -5/+9 $\mu$Hz.

\begin{figure}[t]
\includegraphics[width=9cm]{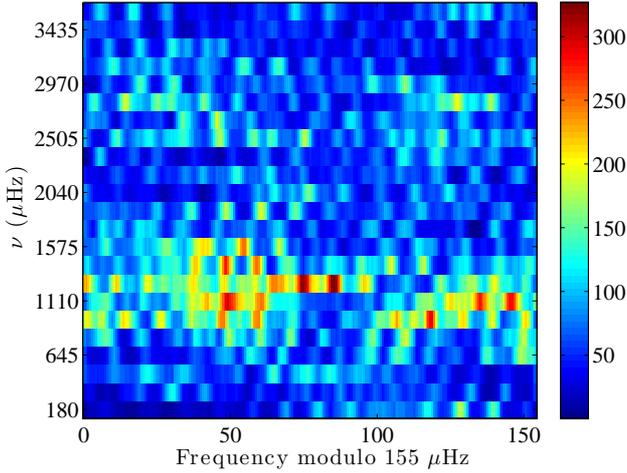}
\caption[]{Echelle diagram of the power spectrum for a folding frequency $\Delta\nu_0  = 155\ \mu$Hz. The power spectrum was divided by the fitted background function and smoothed over five bins, i.e., $3.1 \mu$Hz. It was then divided into consecutive intervals of $155\ \mu$Hz length, which were vertically stacked. The color bar indicates the power in cm$^2$ s$^{-2}$. The existence of two distinct groups of peaks separated by half the folding frequency is compatible with oscillation modes of consecutive degrees (e. g., l = 1 and l = 2).}
\label{fig_3}
\end{figure}

\begin{table}[t]
\caption{Frequencies and amplitudes of the peaks detected by the H0 statistical test with 1\% significance level. }
\begin{tabular}{c c c  c c c }
\hline
$\nu$                       & Velocity                      & Error                            & $\nu$                       & Velocity                     & Error\\
{\tiny $\mu$Hz}      & {\tiny cm s$^{-1}$}    & {\tiny cm s$^{-1}$}     & {\tiny $\mu$Hz}      & {\tiny cm s$^{-1}$}  & {\tiny cm s$^{-1}$}  \\
    \hline
  792   &    44.0   &    -6.2/+3.9   &     1478   &     46.4   &     -6.5/+4.1\\
  854   &    46.7   &    -6.6/+4.2   &     1533   &     37.3   &     -5.3/+3.3\\
  915   &    34.1   &    -4.8/+3.0   &     1615   &     40.9   &     -5.8/+3.7\\
  970   &    48.7   &    -6.9/+4.4   &     1753   &     33.0   &     -4.6/+2.9\\
 1011  &    51.4   &    -7.2/+4.6   &     1939   &     32.0   &     -4.4/+2.8\\
 1066  &    45.7   &    -6.4/+4.1   &     2110   &     30.1   &     -4.2/+2.7\\
 1094  &    42.4   &    -6.0/+3.8   &     2535   &     30.3   &     -4.3/+2.7\\
 1162  &    54.1   &    -7.6/+4.8   &     2714   &     30.6   &     -4.3/+2.7\\
 1245  &    53.8   &    -7.6/+4.8   &     2837   &     36.2   &     -5.1/+3.2\\
 1341   &   51.5   &    -7.3/+4.6   &     2947   &     41.1   &     -5.8/+3.7\\
 1410   &   40.7   &    -5.7/+3.6   &     3071   &     30.7   &     -4.3/+2.7\\

\hline
    \end{tabular}
    \label{table_frequency}
\end{table}
\begin{figure}[t]
\includegraphics[width=9cm]{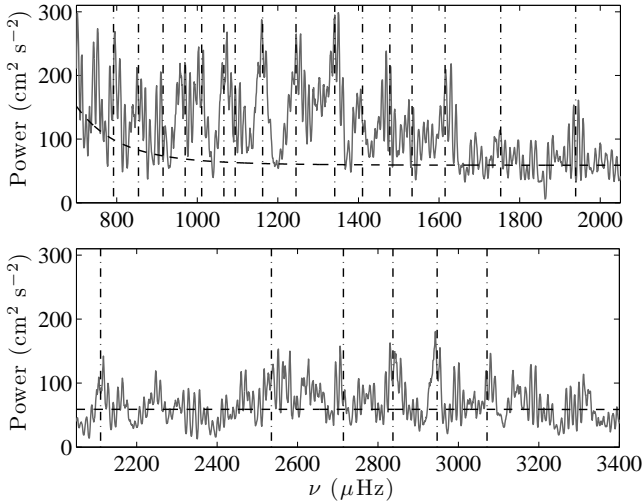}
\caption[]{Top and bottom panels: zoom of the power spectrum between 700 and 3400 $\mu$Hz. The dashed line indicates the background level fit and the vertical dash dotted lines indicate the frequencies of the 22 peaks with 1\% significance level. }
\label{fig_3}
\end{figure}

\section{Evidence of Jovian p-modes}
Our measurements agree with theoretical expectations in terms of the frequency range, the amplitude, and the mean large spacing  \citep{Bercovici_Schubert_1987,Provost_1993} and would correspond to the signature of modes of consecutive degrees (e.g. l=1 and l=2).  Oscillating stars also display similar signatures \citep{Appourchaux_2008_HD49933,Garcia_2009}. However, other effects could a priori cause the measured signal: temperature fluctuations in the interferometer, telescope pointing errors, albedo features on Jupiter, and solar p-modes reflected by Jupiter's atmosphere.

\subsection{The contribution from the instrumental systematics}
\begin{figure}[t]
\includegraphics[width=9cm]{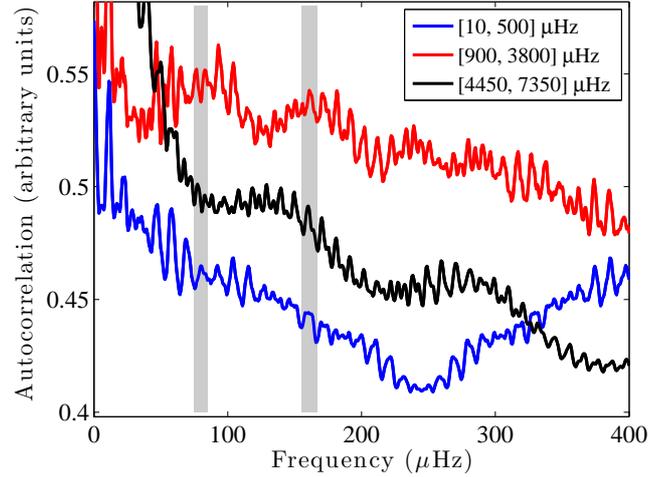}
\caption[]{Autocorrelation of the power spectrum in three frequency ranges. The blue curve shows the low frequencies [10, 500] $\mu$Hz, which are dominated by Jovian rotation, and optical path difference fluctuations of the interferometer, generated by temperature fluctuations. The black curve shows the frequency range dominated by the pointing errors [4450, 7350] $\mu$Hz. The red curve shows the [900, 3800] $\mu$Hz range, where we measure a double excess power. Only this last curve displays a modulation at about $160\pm5\ \mu$Hz, indicated with the grey region, as well as at its half value.}
\label{fig_5}
\end{figure}

Two sources of noise were identified in the data (G08): i) Slow temperature fluctuations of the interferometer cause a spurious signal at frequencies lower than 800 $\mu$Hz. ii) Periodic variations in the position of Jupiter on the detector generate noise whose characteristic frequency is $5920\pm200\ \mu$Hz, i.e., a period of 2.8 min. These two sources of spurious signals could a priori contain harmonics possibly explaining the measured excess power. To assess their contribution to the signal in the frequency range where we detect Jupiter's oscillation modes, in Fig. \ref{fig_5} we compare the autocorrelation of the power spectrum in three frequency ranges: the first [10, 500] $\mu$Hz is dominated by temperature effects, the second [800, 3800] $\mu$Hz is where the power excesses are detected, and the third is [4450, 7350] $\mu$Hz. The autocorrelation of the first frequency range does not have any particular characteristics, as would be expected for a thermal fluctuation spectrum. Its time constant is compatible with the estimated thermal constant time of the Mach-Zehnder interferometer of approximately 30 minutes. It appears that the 6000-$\mu$Hz region is modulated by a low frequency of $136\pm5\ \mu$Hz, which corresponds to a period of about two hours. Such a periodic modulation of the guiding, equal to a fraction of the day, is commonly encountered with telescope guiding systems and may explain the previous incorrect determination of the mean spacing frequency \citep{Mosser_1993,Mosser_2000}. A similar modulation is not observed in the presumed p-mode region. In contrast, we clearly identify two maxima at $80\pm5\  \mu$Hz and $160\pm5\ \mu$Hz, which represents an alternative identification of the large spacing $\Delta\nu_0$. This result excludes an instrumental artifact generating the comb structure.

\subsection{The contribution of solar oscillations}
SYMPA measures the Doppler shift of solar spectral lines after their reflection by Jupiter's cloud decks. Solar oscillations may therefore be detectable in the data, particularly in the secondary excess power, given that solar p-modes extend from 1650~$\mu$Hz to 5000~$\mu$Hz, with a frequency of maximum amplitude at about 3050~$\mu$Hz (e.g.  \citealt{Gelly_2002}). At any given point on Jupiter, the total Doppler shift is the sum of the radial velocities in the solar atmosphere integrated over the entire solar surface, of Jupiter itself locally, and of the relative motions between Jupiter, the Sun, and the observer. However, because we consider the time series associated with antisymmetric modes, all of the global velocity fields (i.e. Jupiter's rotation and signatures from solar oscillations) should cancel out.

To test the possibility of some leakage of the solar modes, it would have been useful to compare the solar and the SYMPA spectra, using the solar full disk Doppler measurements from the GOLF/SOHO instrument. Unfortunately, a four-day interruption in the GOLF data during the SYMPA run prevents us from directly comparing the two time series. BiSON data also suffered a 1-day interruption on April 3 2005 \footnote{http://bison.ph.bham.ac.uk/data.php}. We however computed the spectrum of the GOLF data, for a period of 9.27 days preceding the data interruption, with the same window function. Although some of the peaks match solar frequencies to within $\pm10\ \mu$Hz, this behavior could be serendipitous and does not prove the presence of solar p-modes in the Jovian spectrum. Solar p-modes would be expected to be visible in the symmetric modes, and in particular in the time series corresponding to l=0. However, this time series is affected by guiding and temperature fluctuations and exhibits a higher noise level in the [1000, 3000 $\mu$Hz] frequency range. Solar low-degree p-modes have maximum amplitudes of 20 cm s$^{-1}$ and would be seen with half the sensitivity that we have for Jovian modes. Hence, owing our $\sim$40 cm s$^{-1}$ noise level for this range of frequencies, they are unlikely to explain the excess power measured between 2500 and 3400 $\mu$Hz. In any case, they should not be considered responsible for the main excess power between 800 and 2100 $\mu$Hz.
\begin{figure}[t]
\includegraphics[width=9cm]{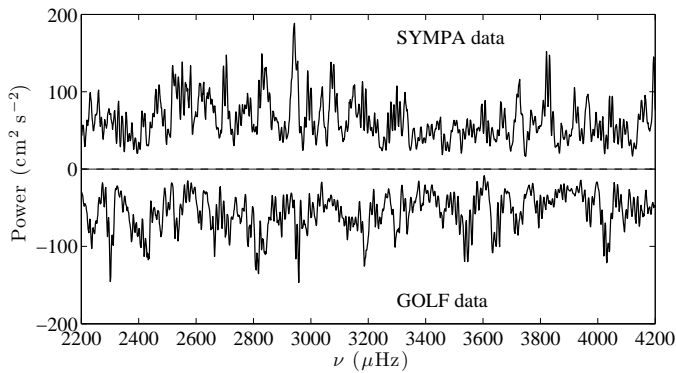}
\caption[]{Power spectra of SYMPA (top) and GOLF/SOHO data (bottom). The Jupiter data is a zoom of the power spectrum plotted in Fig. \ref{fig_1}. The solar spectrum was obtained by using the GOLF/SOHO data on a 9.27-day run, in early April 2005, with the same window function and the same noise level as SYMPA: the GOLF data were interpolated on the same time grid, then a white noise with standard deviation corresponding to those of SYMPA's data (18.6 m s$^{-1}$) was added.}
\label{fig_5bis}
\end{figure}

\subsection{The contribution from Jupiter's meteorological activity}
\begin{figure}[t]
\includegraphics[width=9cm]{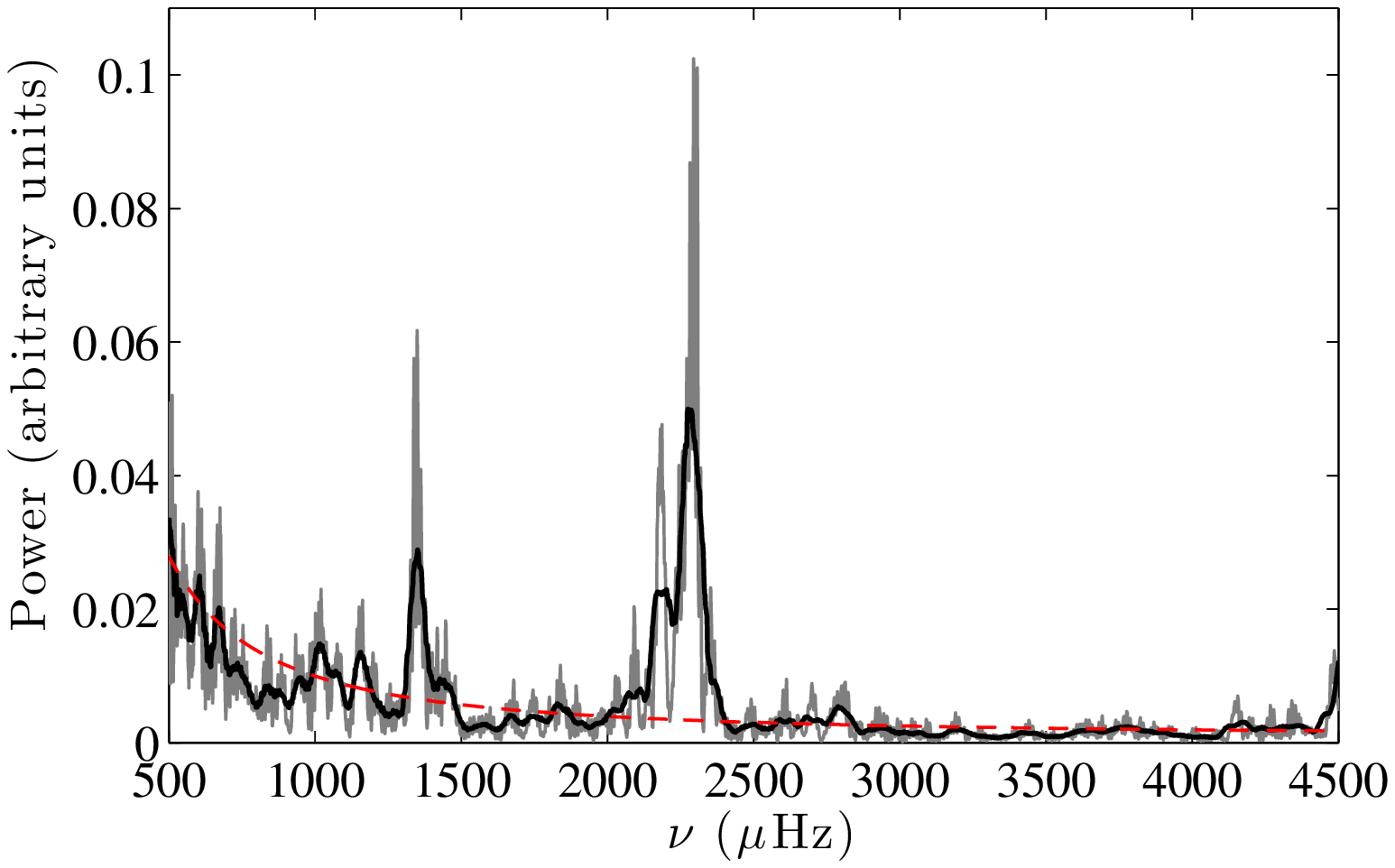}
\includegraphics[width=9cm]{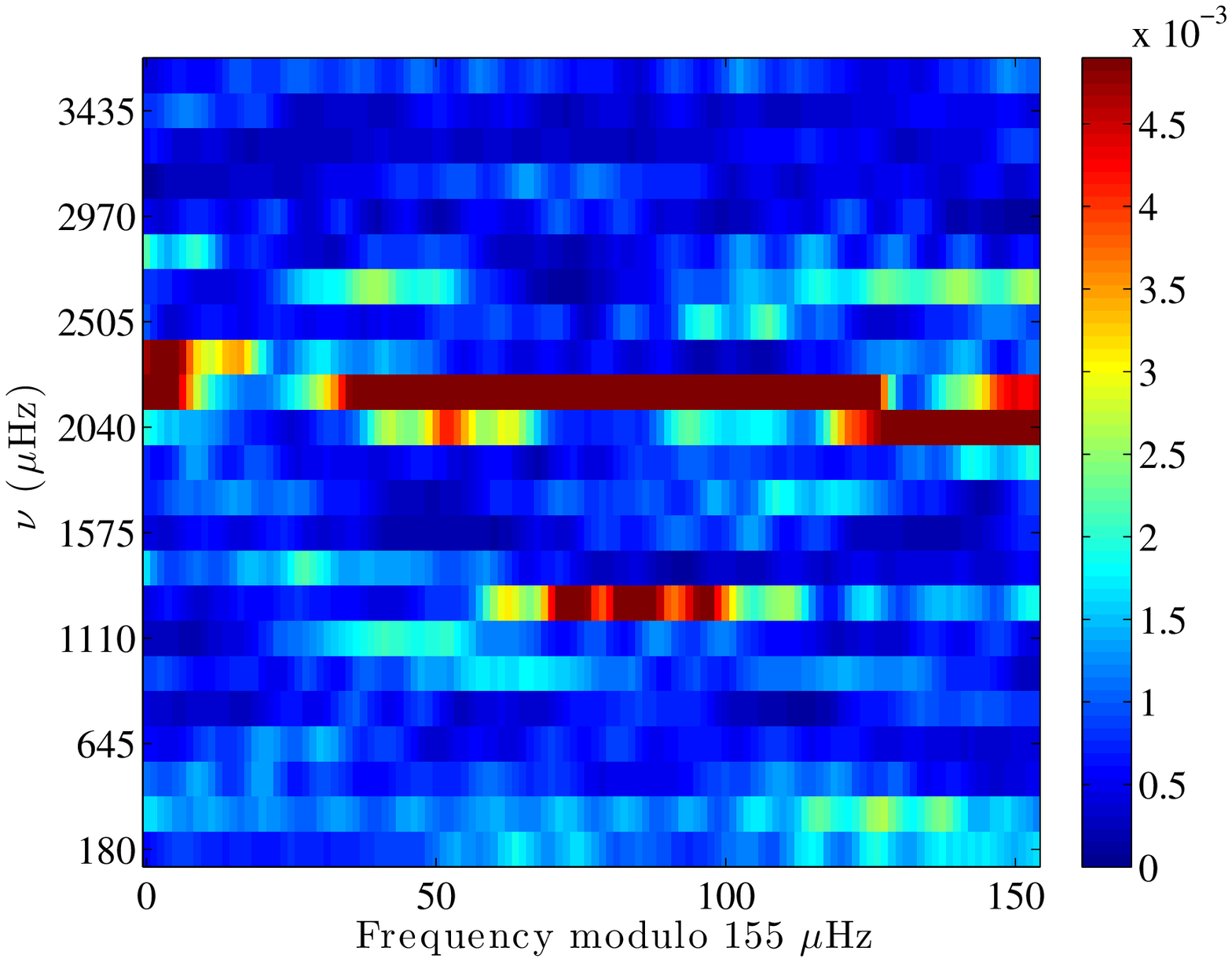}
\caption[]{Top panel: power spectrum of the photometric time series described in ¤2.2. As in Fig. \ref{fig_1}, the thin grey and bold black lines represent the power spectrum smoothed with a 3.1-$\mu$Hz wide triangular window, and a linearly increasing width triangular window, from 14 to 129 $\mu$Hz across the [500, 4500] $\mu$Hz frequency range, respectively. The spectrum is normalized such that its maximum value equals 1. The red dashed line indicates the background level fit. Bottom panel: \'echelle diagram of the power spectrum for a folding frequency of $\Delta\nu_0 = 155\ \mu$Hz, calculated in the same way as in Fig. \ref{fig_3}.}
\label{fig_7}
\end{figure}

Since the detected signal is neither an instrumental nor a solar artifact, it appears to be of Jovian origin, as proven by the presence of twice the Jovian rotation in the comb-like structure (Fig. 2). However, Jupiter is a complex and dynamic planet, with many sources of potentially misleading signals. The inhomogeneities in both the photometric and dynamic structure could indeed affect the measured signal. The effect of quasi-periodic albedo features and its possible contamination of Doppler measurements was studied in \citet{Lederer_1995}. However, they considered the effect of photometric pseudo-periodic features as possible source of spurious velocity signals in the case of spatially unresolved Doppler measurements as in \citet{Schmider_1991} and \citet{Mosser_1993}. Most of the detected frequencies were found below 1000 $\mu$Hz, but that might be a consequence of the limited spatial resolution of their dataset. Unresolved features on Jupiter may indeed affect the Doppler signal and could possibly account for an appreciable amount of the high frequency signal currently identified as p-modes.
These effects should not affect the signal we detect because the velocity map is computed before summation over all pixels and photometric contribution is thus cancelled out. As we simultaneously recorded the photometric images, it is possible to compute the power spectrum of the photometric time series associated with our measurements. For each exposure, the sum of the four outputs of the instrument is an image of Jupiter with the fringe pattern suppressed \citep{Schmider_2007}. We multiplied each image for the whole run with the pair of masks l,m = 1,0 and l,m = 1,1, eliminating the outer 25\% of Jupiter's diameter, to obtain two new times series. The power spectrum of the mean time series was then calculated with the discrete Fourier transform. Neither the power spectrum nor the \'echelle diagram associated with $\Delta\nu_0$ present any similarity to the power spectrum of the radial velocity time series (Fig. \ref{fig_7}). As expected, the peaks due to albedo features are integer multiples (48, 78, 82, for the most important) of the Jovian rotation frequency, corresponding to spatial pseudo-periodic features on Jupiter.
The main peak in the photometric power spectrum ($\sim2250\ \mu$Hz) corresponds to a range of minimum energy in the velocity spectrum. This proves that our processing fully suppresses the photometric fluctuations in the velocity measurement. Moreover, in the photometric signal, the absence of the comb-like structure detected in the velocity spectrum proves that the velocity signal cannot be associated with the Jovian dynamical inhomogeneities that are associated with the pseudo-periodic albedo features. This is unsurprising as these dynamical patterns have mainly horizontal components and would contribute to the Doppler measurement only at the limb, which was removed during the data processing. We therefore conclude that acoustic eigenmodes are the most probable explanation of the measured signal.

\begin{figure}
\includegraphics[width=9cm]{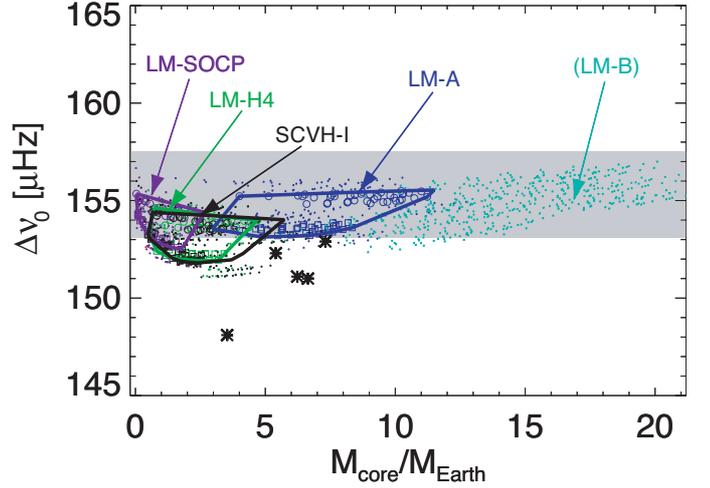}
\caption[]{Masses of the central core and values of  $\Delta\nu_0$ predicted by interior models of Jupiter. Colored symbols represent solutions obtained for different EOSs of hydrogen (as labeled) and using consistent EOSs for helium, ices, and rocks (see \citet{Saumon_guillot_2004} for a full description of the models). Boxes, open squares, and open circles correspond to solutions matching all available constraints. Dots correspond to models that fail to match the observed fourth order gravitational moment J4. The models assume a two-layer hydrogen-helium envelope overlying a core made either of pure rocks (circles) or pure ices (squares). Asterisks correspond to older model solutions in which ices and rocks were modeled using the helium EOS \citep{Gudkova_1995}. The grey area indicates the 1-$\sigma$ constraint obtained from our determination of the large separation of the modes.}
\label{fig_8}
\end{figure}

\subsection{Comparison with Jupiter's model}
The inferred $\Delta\nu_0$ value agrees with the predictions of modern models for the interior of Jupiter \citep{Saumon_guillot_2004}. This result is independent of the equation of state of hydrogen used for the modeling and even extends to models that do not fit Jupiter's J4 measurement (Fig. \ref{fig_8}). Older interior models that used an equivalent helium fraction to mimic the presence of heavy elements \citep{Gudkova_1995} yield values of $\Delta\nu_0$ that are slightly too small, thus showing that the present determination is already useful for validating Jupiter models. The theoretical $\Delta\nu_0$ value depends only weakly on factors such as the core mass, but we note that the core composition parameter yields a rather significant $\sim1\mu$Hz change when varied between pure rock and pure ice models, which slightly favors the former solutions. More accurate determinations of $\Delta\nu_0$ can thus provide with information about Jupiter's central core, a crucial element to understand the planet's formation. However, the tightest constraints will be provided by the comparison of observed with theoretical individual mode frequencies.

\section{Discussion and conclusion}
\subsection{Detection of Jovian p-modes between 800 and 2100 $\mu$Hz}
We applied the most sophisticated analysis methods of asteroseismology to reanalyze the Jovian data, obtained by the SYMPA instrument, in April 2005, at Teide observatory. This led to the detection of a clear excess power between 800 and 2100 $\mu$Hz and a secondary excess power between 2400 and 3400 $\mu$Hz, which are both modulated by a comb-like structure. We could easily rule out the temperature and pointing effects, and albedo features on Jupiter as causing the measured features in the power spectrum of the photometric time series associated with the dataset. Solar-p modes are similarly unable to account for the main excess power. 

In terms of the secondary excess power, the short observing time associated with a low duty cycle, prevents us from being able to explain such a double bump in terms of the three following hypotheses: 1) It is a snapshot of the modes that were stochastically excited in early April 2005. 2) It reveals a variation in the mode amplitude as a function of frequency. This hypothesis cannot be excluded in the case of Jupiter, as the inner structure could contain several discontinuities that could affect the quality factor of modes with different radial order.  3) The secondary excess power is the reflection of solar modes. The last point is the least probable because the solar large-spacing signature does not appear in our data. Moreover, since the cut-off frequency of Jovian modes is about 3500 $\mu$Hz \citep{Lognonne_Johnson_2007}, finding modes at these frequencies would not be surprising. Finally, the secondary excess power also mirrors the double ridge structure that is observed in the main excess power in the \'echelle diagram (Fig. \ref{fig_3}).  

In any case, the most probable hypothesis is that the signal between 800 and 2100 $\mu$Hz is generated by Jupiter's acoustic modes. We measured the global seismic parameters, i. e., the frequency of maximum amplitude at $\nu\ind{max}=1213\pm50\ \mu$Hz and the mean large spacing $\Delta\nu_0=155.3\pm2.2\ \mu$Hz. The large spacing value agrees with theoretical models of the internal structure of Jupiter. In addition, \citet{Mosser_1995} calculated the quality factor of the Jupiter cavity, then infered that the mode lifetime is longer than 12 days. Thus, the mode damping is almost negligible during a nine-day run and we estimate the mode amplitude using the pure sine assumption, which causes the maximum mode amplitude to be $49_{-10}^{+8}$~cm s$^{-1}$. The SNR of the detection is similar to the first asteroseismic results for solar-like targets, from data acquired by the CoRoT satellite \citep{Baglin_2002}, in particular those obtained for HD~181906 \citep{Garcia_2009}.

The large error bar in the peak frequencies ($-5/+9\ \mu$Hz) prevents us from identifying the modes. From Table \ref{tab_1}, the leakage between the masks implies that the detected peaks could correspond to modes of degree (l = 1, m =$\pm1$), (l = 2, m=$-1,-2$), (l = 3, m=$0, 1, 2, 3$), and (l = 4, m=$-4, -3, -2,-1$), which are almost uniformly distributed as a function of frequency, by considering that the Jovian rotation introduces a frequency splitting of 28~$\mu$Hz and that modes of consecutive degrees are spaced by about half the large spacing.

\subsection{A new start for giant planet seismology}
To confirm the present observations, further observations are planned with an improved model of SYMPA, including an accurate pointing system, for the 2012 and 2013 Jovian oppositions. However, seismic measurements will strongly constrain Jupiter's inner density profile only with data of a significantly higher SNR, when modes are unambiguously identified in the spherical harmonics base. The main limitation of ground-based measurement arises from the poor duty cycle and atmospheric turbulence. Obviously, the optimal observing conditions would be obtained from space observations, which is the purpose of the Doppler spectro-imager DSI/ECHOES \citep{Schmider_2010} proposed to the ESA Laplace mission \citep{Blanc_2009}.

In any case, our results provide the clearest evidence yet of Jovian global oscillations. These measurements open the possibility of directly probing the planet's interior and determine its global composition, core size, presence of phase transitions, and even internal rotation. The technique should be applicable to all giant planets in the Solar System, with unique prospects for the understanding of their composition and formation.

\begin{acknowledgements} The authors thank Beno\^{\i}t Mosser, s.a.s. Thierry Appourchaux, Philippe Lognonn\'e, and John W. Leibacher for useful comments, and Per\'e Pall\'e and Manuel Alvarez for their help in the observations. They are grateful to the Instituto de Astrof'sica de Canarias, the OPTICON Network, the MEO/Calern telescope team, the San Pedro Martir Observatory for the observing time. The SYMPA project was funded by the French CNRS, INSU, the Minist\`ere de l'Enseignement Sup\'erieur et de la Recherche, the Observatoire de la C\^ote d'Azur.
\end{acknowledgements}
\bibliographystyle{aa}
\bibliography{bibi}

\end{document}